\documentclass[12pt, a4paper]{article}
\usepackage[dvips]{graphics}

\title
{Causality, delocalization and positivity of energy.}

\author{E. Karpov$^\ast$,  G. Ordonez$^{\ast\ast}\,^\ast$, T. Petrosky$^{\ast\ast}\,^\ast$
I. Prigogine$^\ast\,^{\ast\ast}$\\ and G. Pronko$^{\ast\ast\ast}\,^\ast$\\
\begin{tabular}{l l}
$^\ast$   & {\small \it International Solvay Institutes for Physics and Chemistry,} \\
          & {\small \it C.P. 231, Campus Plaine ULB, Bd. du Triomphe, Bruxelles 1050, Belgium} \\
$^{\ast\ast}$ & {\small \it Center for Studies in Statistical Mechanics and Complex Systems,} \\
              & {\small \it The University of Texas at Austin, Austin, Texas 78712, USA }\\
$^{\ast\ast\ast}$ & {\small \it Institute for High Energy Physics, 
                     Protvino, Moscow region 142284, Russia} \\
\end{tabular}}
\date{}
\begin{document}
\maketitle

\abstract{In a series of interesting papers
G. C. Hegerfeldt has shown that 
quantum systems with positive energy initially localized in a finite region, 
immediately develop infinite tails.
In our paper Hegerfeldt's theorem is analysed using quantum and classical wave packets.
We show that Hegerfeldt's conclusion remains valid in classical physics.
No violation of Einstein's causality is ever involved. 
Using only positive frequencies, complex wave packets are constructed 
which at $t = 0$ are real and finitely localized and which, furthemore, 
are superpositions of two nonlocal wave packets.
The nonlocality is initially cancelled by destructive interference. However this cancellation
becomes incomplete at arbitrary times immediately afterwards. 
In agreement with relativity the two nonlocal wave packets 
move with the velocity of light, in opposite directions.}

\newpage
\section{Introduction}

Are there deviations from Einstein's causality? G. C. Hegerfeldt has written 
\cite{Heg98}:
``Positivity of the Hamiltonian alone is used to show that particles, 
if initially localized in a finite region, immediately develop infinite tails.''
This seems to imply superluminality. One of his examples is 
the Fermi problem \cite{Fermi} of two atoms coupled by a radiation field. 
Consider the initial condition when one of the atoms is in an excited state, 
the other in the ground state, and no photons are present.
The probability to find the second atom in an excited state is non-vanishing immediately
after the initial moment, 
independently of the distance between the atoms \cite{Heg98}, \cite{Heg74} - \cite{Heg94}. 
Hegerfeldt's arguments are based on the analyticity of the
expectation values of the operator $N(V)$,
which gives the probability to find a particle inside a finite volume $V$.
He showed that a state in a quantum system with positive energy 
localized in a finite volume $V$ at the instant $t=0$,
will develop infinite tails immediately afterwards.
Positivity of energy plays an essential role in his proof.
In this paper we present an illustration of Hegerfeldt's theorem,
without any appeal to superluminality.
We apply Hegerfeldt's consideration to wave packets. Moreover, we show
that Hegerfeldt's effect appears even for classical fields, if wave packets are
constructed from positive frequencies (corresponding positive energy quantum fields).

We first study the positive-frequency solutions of the classical wave equation (section 2). 
We consider wave packets $\Phi(x,t)$ localized at $t=0$.
The localization is due to interference of the two complex solutions, 
each propagating causally
\equation \label{twp}
  \Phi(x,t) 
    =\psi(x-t) + \psi^*(x+t) 
\endequation
where ``$^*$'' denotes complex conjugation and we take $c=1$. We show that
both these wave packets are delocalized. They present long tails, extending to
arbitrary distances and decaying according to a power law.
As we shall show, the ``nonlocal effect'' can also be understood from the point
of view of the initial conditions. Indeed, in our construction of the solution 
(\ref{twp}) we shall use two conditions; one is the initial condition 
of the local shape of the field $\Phi(x,t)$ and the other is the condition 
of the positivity of frequencies. The frequency positivity
replaces the usual initial condition on the time derivative of the field 
$\partial\Phi/\partial t$. We shall show that our initial condition 
with positive frequencies leads to the nonlocality of $\partial\Phi/\partial t$
at $t=0$.

In section 3 we show that similar conclusions are obtained for the wave packet 
of a free field in relativistic quantum field theory. We construct an operator
reminding of the position operator of Newton-Wigner \cite{NW}. The expectation
value of this operator with the state corresponding to our wave packet is
local at $t=0$. However, it has infinite tails which are ``hidden'' at time $t=0$, 
but emerge immediately afterwards.
We may call this effect a ``curtain'' effect. No superluminal propagation is
involved.
We note that 
at the same time other quantities such as the energy density have a
nonlocal expectation value in the same state even for $t=0$. 
It should be also pointed out that for the Dirac equation 
there are no positive energy solutions which can be localized in a 
finite region (see \cite{Heg98}).
This demonstrates
that localization in relativistic quantum field theory cannot be ``complete''.

\section{Classical wave packets}
Consider classical wave packets constructed by the
solutions of the wave equation with positive frequency and localized at time $t=0$.
We show that these wave packets will spread immediately over the whole space. 
Curiously we have not found any reference to this effect in the literature.
We start from the wave equation on the real line ($c=1$).
\equation \label{we}
  \left(\frac{\partial^2}{\partial t^2} 
      - \frac{\partial^2}{\partial x^2}
  \right)\Phi(x,t) = 0
\endequation
The general {\it complex } solution of Eq. (\ref{we})
is, by the Fourier transform, of the form
\equation \label{gensol}
  \Phi(x,t) = \frac{1}{2\pi}\int^{\infty}_{-\infty}~dk
              \left\{\phi_+(k)e^{-i\omega_k t} + \phi_-(k)e^{i\omega_k t}\right\}
              e^{ikx}
\endequation
where $\omega_k  =  |k|$ and where $\phi_\pm(k)$ are arbitrary functions.
To determine $\phi_+$ and $\phi_-$ one can use the two initial conditions
$\Phi(x,0)$ and $\dot{\Phi}(x,0)$. However, one can also consider 
the special class of positive-frequency solutions to Eq. (\ref{we}),
i.e. $\phi_-(k) \equiv 0$ and
\equation \label{pes} 
  \Phi_+(x,t)=\frac{1}{2\pi}\int^{\infty}_{-\infty}~dk
              \phi_+(k)e^{-i\omega_k t}e^{ikx}
 \endequation
These positive-frequency solutions are determined by the 
initial condition $\Phi(x,0)$. 
Note that relation (\ref{pes}) leads to a complex field for $t\ne 0$, 
even if $\Phi_+(x,0)$ or $\phi_+(k)$ are real.
Consider as an example a localized (rectangular) wave packet with center $x_0$,
and width $2b$ at time $t=0$:
\equation \label{phi0}
  \Phi_{x_0,b}(0) = \frac{1}{2b}\Theta(b-|x-x_0|)
\endequation
The normalization has been chosen so that the integral of this function over $x$
is equal to one. Then the function $\phi(k)$ is:
\equation \label{phik}
  \phi_+(k) = \frac{1}{2b}\int_{-\infty}^{+\infty}\!\!\!dx\,e^{-ikx}\Theta(b-|x-x_0|)
\endequation
where $\Theta(x)$ is the step function,
which is $0$ for $x$ negative, and $1$ for $x$ positive.
Then the function $\Phi(x,t)$ in (\ref{pes}) is given by
\equation \label{fou1}
  \Phi(x,t) 
    = \frac{1}{4\pi b}\int_{-\infty}^{+\infty}\!\!\!dk
                    \int_{x_0-b}^{x_0+b}\!\!\!dx'\,e^{-i|k|t + ik(x-x')}                   
\endequation
This is a sum of two functions corresponding to two wave packets
moving in opposite directions,
\equation \label{fou2}
  \Phi(x,t) = \psi(x-t) + \psi^*(x+t)
\endequation
where
\equation \label{psix}
  \psi(x) = \frac{1}{4\pi b}\int_{x_0-b}^{x_0+b}\!\!\!dx'
                          \int_{0}^{+\infty}\!\!\!dk\,e^{ik(x-x')}  
\endequation
To evaluate the integral over $k$ we introduce the usual regularisation 
by adding a positive infinitesimal to $x$, which leads to
\equation \label{psixe}
  \psi(x)    = - \frac{1}{4\pi bi}
        \int_{x_0-b}^{x_0+b}\!\!\!\frac{dx'}{x-x'+i0}
\endequation
After integration over $x'$ we obtain:
\equation \label{psixint}
  \psi(x)
    = \frac{i}{4\pi b}[\ln(x-x_0+b+i0)-\ln(x-x_0-b+i0)]
\endequation
The logarithm of a complex number is given by
\equation \label{ln}
  \ln(z) = \ln|z| + i(\arg(z)+2\pi n)
\endequation
where $n$ is an integer. In order to have both terms in (\ref{psixint})
on the same branch of the logarithm 
we take $n=0$ for both of them 
(due to the difference of the two terms in (\ref{psixint})
the result does not depend on the particular value of $n$). 
The argument of $x+i0$ can be expressed as
\equation \label{arg}
  \arg(x+i0)=\frac{\pi}{2}(1-\mathrm{sign}(x))
\endequation
where $\mathrm{sign}(x)=x/|x|$ is the sign of $x$.
Then, inserting (\ref{ln}) and (\ref{arg}) into (\ref{psixint}) we obtain
\equation \label{psixx}
  \psi(x)
     = \frac{1}{8b}\left(\mathrm{sign}(x-x_0+b) - \mathrm{sign}(x-x_0-b)\right)
     + \frac{i}{4\pi b}\ln{\left|\frac{x-x_0+b}{x-x_0-b}\right|}
\endequation
We see that the function $\psi(x)$ in (\ref{psixx}) consists of a local real part
($\mathrm{sign}$) and a nonlocal imaginary part ($\log$). 
For $t \neq 0$ it is sufficient to replace $x$ by $x-t$ in (\ref{fou2}). Similar result is obtained for
$\psi^*(x+t)$.
As a result, the function $\Phi(x,t)$
is also nonlocal because it is the superposition 
of the two complex functions $\psi(x-t)$ and $\psi^*(x+t)$ in (\ref{fou2}), 
which describe nonlocal objects moving with the speed of light 
in opposite directions.
However, at $t=0$ the imaginary parts cancel each other (see Fig. 1),
and we recover our localized initial condition (\ref{phi0}), 
because only the real parts of these functions, which are local, remain.
In all our figures time $t$ is measured in seconds ($s$), the coordinate $x$ is measured
in ``light seconds'' ($ls$) and wave packet amplitudes are dimensionless.
Fig. 2 corresponds to $t=0.25\, s$. At this moment the real local parts of 
$\psi(x-t)$ and $\psi^*(x+t)$ have moved in opposite directions. 
The nonlocal imaginary parts of $\psi(x-t)$ and $\psi^*(x+t)$ have also
shifted in opposite directions and no more cancel each other.
At this time, the two waves overlap and we have
\equation \label{overlap}
  |\Phi(x,t)|
    \equiv|\psi(x-t)+\psi^*(x+t)| 
    \neq |\psi(x-t)| + |\psi^*(x+t)|
\endequation
At $t=1\,s$ in Fig. 3 the overlapping is small and we have
\equation \label{noverlap}
  |\Phi(x,t)| 
    \approx |\psi(x-t)| + |\psi^*(x+t)|
\endequation
We see that the initial condition $\Phi(x,0)$ is local (Fig. 1)
only because at $t=0$ the nonlocal parts cancel each other completely
by destructive interference. We may describe the appearance of nonlocality
as a sort of ``curtain effect''. The nonlocal nature of each wave packet
$\psi(x-t)$ and $\psi^*(x+t)$ is hidden behind a ``curtain'' at the initial time 
and emerges immediately afterwards. Each of the nonlocal wave packets is complex
and propagates at the speed of light.

In conclusion, we have illustrated Hegerfeldt's theorem 
for classical wave packets. We see that the localization of
wave packets corresponding to positive frequency is unstable and
involves complex space structures.

Note that the localization of a positive frequency wave packet
is not ``complete'', because the time derivative of the function $\Phi(x,t)$ 
is nonlocal even at $t=0$:
\equation \label{psider}
  \left[\frac{\partial\Phi(x,t)}{\partial t}\right]_{t=0}
    = \frac{i}{2\pi b}\left(\frac{1}{x-x_0-b} - \frac{1}{x-x_0+b}\right)
\endequation
The wave equation being of second order demands two initial conditions: 
for the function itself and for its time derivative.
The additional requirement of positivity of frequency
replaces the second condition. 
There are no wave packets containing only positive frequency modes,
which are localized together with their time derivative \cite{Pavlov}.

\section{Relativistic quantum field}

We turn now to relativistic quantum field theory. 
We show that the previous discussion is applicable to relativistic 
quantum particles. In this case the condition $\omega_k>0$ appears naturally 
since the energy $E=\hbar \omega_k$ must be positive (we take $\hbar=1$). 
We consider massless particles with no spin (``photons'').
To simplify our consideration we use again a 1+1-dimensional spacetime.
In terms of second quantization we have the scalar field operator 
\equation \label{psifield}
  \hat{\psi}(x,t) = \int_{-\infty}^{+\infty}\!\!\!d\tilde{k}
            \left(a^\dag_k e^{i(\omega_k t - kx)}
                + a_k e^{-i(\omega_k t - kx)}
            \right)
\endequation
where $d\tilde{k} = dk/(4\pi \omega_k)$ is a relativistic invariant measure and
\equation \label{drq}
  \omega_k =|k|
\endequation
with  $c=1$. 
The creation and annihilation operators $a^\dag_k$ and $a_k$ 
of the photon with wave vector $k$ obey the commutation relation
\equation \label{cr}
  \left[a^\dag_k, a_{k'}\right] = 4\pi \omega_k \delta(k-k')
\endequation
We construct a wave packet from a linear combination of normal modes.
\equation \label{stphi}
  |\Phi_{x_0,b}(t)\rangle
    = \int_{-\infty}^{+\infty}\!\!\!d\tilde{k}\,
      \phi_{x_0,b}(k)e^{-i\omega_kt}a^\dag_k|0\rangle
\endequation
Here $|0\rangle$ is the vacuum state for the field (\ref{psifield}). 
The fact that our wave packet is obtained by the action of creation operators
on the vacuum state implies that the state consists of normal modes with
positive energy. As before, we chose the function $\phi_{x_0,b}(k)$
so that the wave packet is localized 
at the time $t=0$ in a domain with center $x_0$ and width $2b$,
\equation \label{kphi}
  \phi_{x_0,b}(k) 
    = (2\omega_k)^{1/2}\int_{-\infty}^{+\infty}\!\!\!dx\, e^{ikx}\Phi_{x_0,b}(x,0)
\endequation
where  
\equation \label{xphi}
  \Phi_{x_0,b}(x,0)=\frac{1}{(2b)^{1/2}}\Theta(b-|x-x_0|)
\endequation
The  function $\Phi_{x_0,b}(x,0)$ is normalized to ensure the normalization of 
the state  $|\Phi_{x_0,b}(0)\rangle$ in (\ref{stphi}). 

Let us introduce the operator $\rho(x)$:
\equation \label{ro}
  \rho(x)=a^\dag(x)a(x)
\endequation
where $a^\dag(x)$ and $a(x)$ are defined by
\begin{eqnarray}
  a^\dag(x) 
    & = & \int_{-\infty}^{+\infty}\!\!\!d\tilde{k}(2\omega_k)^{1/2}\,e^{-ikx}a^\dag_k 
\nonumber \\
\label{adx}
  a(x) 
    & = & \int_{-\infty}^{+\infty}\!\!\!d\tilde{k}(2\omega_k)^{1/2}\,e^{ikx}a_k 
\label{ax}
\end{eqnarray}
These operators satisfy the commutation relation
\equation \label{xcommutaion}
  [a(x), a^\dag(x')] = \delta(x-x')
\endequation
This construction follows the ideas of the construction of positions operators 
by Newton and Wigner \cite{NW}.
We shall express the localization of our state $ |\Phi_{x_0,b}(t)\rangle$
in terms of the expectation value of the operator $\rho(x)$.
We call a state localized if the expectation value of $\rho(x)$ in this state 
vanishes when $x$ is outside a finite region.
Our choice of  $\phi_{x_0,b}(k)$ in (\ref{kphi}) and $\Phi_{x_0,b}(x)$ in (\ref{xphi})
guarantees that the state $ |\Phi_{x_0,b}(t)\rangle$ 
is localized at $t=0$ in the domain $[x_0-b,x_0+b]$, i.e.,
\equation \label{localization}
  \langle \Phi_{x_0,b}(0)|\rho(x)|\Phi_{x_0,b}(0)\rangle
    = \frac{1}{2b}\Theta(b-|x-x_0|)
\endequation
Using (\ref{adx}) we obtain the time evolution of this quantity
\equation \label{prp}
  \langle \Phi_{x_0,b}(t)|\rho(x)|\Phi_{x_0,b}(t)\rangle
    = \int\!\!\!\int_{\-\infty}^{+\infty}\!\!\!d\tilde{k}\,d\tilde{k}'\,
      (4\omega_k\omega_{k'})^{1/2}e^{-i(k-k')x}
      \langle \Phi_{x_0,b}(t)|a^\dag_k a_{k'}|\Phi_{x_0,b}(t)\rangle
\endequation
where $\langle \Phi_{x_0,b}(t)|a^\dag_k a_{k'}|\Phi_{x_0,b}(t)\rangle$
is expressed using our form of the wave packet (\ref{stphi})
as follows:
\equation \label{paap}
  \langle \Phi_{x_0,b}(t)|a^\dag_k a_{k'}|\Phi_{x_0,b}(t)\rangle
    = \int \!\!\!\int_{\-\infty}^{+\infty}\!\!\!d\tilde{l}\,d\tilde{l}'\,
      \phi^*_{x_0,b}(l)\phi_{x_0,b}(l')e^{i(\omega_l-\omega_l')t}
      \langle 0|a_{l}a^\dag_k a_{k'}a^\dag_{l'}|0\rangle
\endequation
Using the commutation relation (\ref{cr}) we integrate (\ref{paap}) 
over $l$ and $l'$ and then insert the result into (\ref{prp}).
Taking into account the positivity of energy (\ref{drq})
and the form of the function $\phi_{x_0,b}(k)$ in (\ref{kphi}) we obtain
\equation \label{paap2}
  \langle \Phi_{x_0,b}(t)|\rho(x)|\Phi_{x_0,b}(t)\rangle
   = \left| \frac{1}{2\pi (2b)^{1/2}}\int_{x_0+b}^{x_0-b}\!\!\!dx'\,
            \int_{-\infty}^{+\infty}\!\!\!dk\, e^{-i|k|t+ik(x-x')}
      \right|^2
\endequation
By comparison with (\ref{fou1}) we see that this quantity is equal to 
the absolute value squared of the classical function $\Phi(x,t)$ 
up to the normalization constant.
Our discussion of nonlocality remains, therefore, also valid in the quantum case,
and the expression inside the absolute value in  (\ref{paap2}) is a superposition of two 
nonlocal wave packets that move in opposite directions at the speed of light.
In the appendix give a second example using an analogy with Fermi's problem 
\cite{Fermi}.

Let us note that in quantum field theory localization depends on the observable.
If a state is local from the point of view of one observable, 
it can be nonlocal from the point of view of another.
In our example the expectation value of the operator $\rho(x)$ is local in the 
state $|\Phi_{x_0,b}(0)\rangle$. At the same time,
the energy density of the field in the same state is nonlocal. 
Indeed, the energy density $T_{00}(x)$ of the free massless field
(\ref{psifield}) is 
\equation \label{enden}
  T_{00}(x)
    =\frac{1}{2}\left(\left(\frac{\partial\hat\psi}{\partial t}\right)^2
                    + \left(\frac{\partial\hat\psi}{\partial x}\right)^2
                \right)
\endequation
It contains the derivatives of the field operator. (As we have seen in classical case,
the time derivative of the function $\Phi(x,t)$ is nonlocal even at $t=0$).
To determine the expectation value of $T_{00}(x)$ 
in the state $|\Phi_{x_0,b}(0)\rangle$
taking into account the positivity of energy (\ref{drq}) we first
calculate this expectation value for a finite $t$ 
and then take the limit $t \rightarrow 0$.
Using (\ref{psifield})-(\ref{xphi}) obtain:
\equation \label{energy}
  \langle \Phi_{x_0,b}(0)|T_{00}(x)|\Phi_{x_0,b}(0)\rangle
    =  \frac{1}{4\pi b}\left(\frac{1}{|x-b|}+\frac{1}{|x+b|}
                          -  \frac{1-\mathrm{sign}(x-b)\,\mathrm{sign}(x+b)}
                                  {\sqrt{|x-b|}\sqrt{|x+b|}}
                       \right)
\endequation
where we put $x_0=0$ to simplify the expression. 
This quantity is obviously nonlocal. 

\section{Conclusion}
Positivity of energy for quantum fields (or frequency for classical fields) leads to a decomposition of
localized wave packets in terms of nonlocal wave packets  with long tails. The long tails,  which cancel
each other initially, appear immediately afterwards as the nonlocal wavepackets move in opposite directions.
In our examples the long tails decay with the distance $x$ according to $b/x$ for $b/x \ll 1$, 
where $b$ is the size of the localized wave packet.
They are precursors to the usual wave propagation. Although we may have instant
interactions, these are not the result of  superluminal propagation, but of ``preformed'' structures. 

We shall study the interaction between nonlocal structures in a separate paper. 
We have then ``contact interactions'', due to the overlapping of the long tails.
We shall also show that the photon clouds around atoms and molecules are nonlocal,
which leads to the precursor effect and eliminates the apparent deviation 
from causality in Fermi's two-atom problem.
However, it is true that the two atoms ``feel'' each other instantaneously.
Even inside a relativistic theory (the wave equation is Lorenz invariant) there is
place for instantaneous interactions due to nonlocality.

Einstein's relativistic events are associated to four dimensional points.
Here we see nonlocal but still relativistic events that are due to the instability 
of localization as  shown in the examples presented in this paper.

\vskip 1cm

\begin{flushleft}
{\bf ACKNOWLEDGEMENTS}
\end{flushleft}
The authors would like to thank Prof. B. Pavlov and Prof. I. Antoniou
for comments and fruitful discussions.
This work was carried out with financial support of the
International Solvay Institutes, the European Commission ESPRIT Project 28890 NTGONGS.
This work was partially supported by the Engineering Research Program of
the Office of Basic Energy Sciences at the Department of Energy Grant No.
DE-FG03-94ER14465, the Robert A Welch Foundation Grant No. F-0365.
\vskip 1cm

\begin{flushleft}
{\bf APPENDIX}
\end{flushleft}
\newcounter{equationA}
Here we show an analogy of our problem with Fermi's two-atom problem.
We prepare our wave packet $|\Phi_{x_0,b}(t)\rangle$
``localized'' at $t=0$. At time $t$
we project this state on an second wave packet 
$|\Phi_{x_1,b}(0)\rangle$, which is
localized in a  domain with center $x_1 \neq x_0$ and width $2b$, 
and plays the role of a measurement device.
We choose $x_1$ so that $  x_1-x_0 > 2b$. Then, at $t=0$ these two states do not overlap
and the scalar product $\langle\Phi_{x_1,b}(0)|\Phi_{x_0,b}(t)\rangle$ vanishes.
We consider this scalar product as a function of $t$, i.e., at each moment $t$ 
we project  the evolving wave packet $|\Phi_{x_0,b}(x,t)\rangle$
on the localized state $|\Phi_{x_1,b}(0)\rangle$.
As we have shown, the initially localized packet, which evolves in time,
is nonlocal immediately after $t=0$ and our scalar product has a non-vanishing value.
This can be interpreted as the second (localized) wave packet ``feeling'' 
the existence of the first one even at $t < x_1-x_0-2b$
when the causal component of the first wave packet still 
did not reach the domain of localization of the second wave packet.
The scalar product  $\langle\Phi_{x_1,b}(0)|\Phi_{x_0,b}(t)\rangle$
grows as the overlapping of the two wave packets increases (see Fig. 4). 
We expect some essential change of this growth
when the main part of the first wave packet corresponding to the position of its
local (``causal'') component reaches the 
domain of localization of the second wave packet.
Then, as the moving component goes away, the scalar product decreases.
To show this, we write this scalar product using (\ref{stphi}) 
in the following form
$$
  \langle\Phi_{x_1,b}(0)|\Phi_{x_0,b}(t)\rangle
    = \int\!\!\!\int_{-\infty}^{+\infty}\!\!\!dk\,dk'
      \phi^*_{x_1,b}(k)\phi_{x_0,b}(k')e^{-i\omega_kt}
      \langle 0|a_k a^\dag_{k'}|0\rangle
\refstepcounter{equationA}
\eqno{(A\theequationA)}
\label{PPaa}
$$
We perform the integration over $k'$ 
with the help of the commutation relation (\ref{cr}).
Then, inserting (\ref{kphi}) and (\ref{xphi}) we obtain
$$ 
  \langle\Phi_{x_1,b}(x,0)|\Phi_{x_0,b}(x,t)\rangle
    =\frac{1}{4\pi b}\int_{x_1-b}^{x_1+b}\!\!\!dx'\int_{x_0-b}^{x_0+b}\!\!\!dx''
     \int_{-\infty}^{+\infty}\!\!\!dk\,e^{-i|k|t+ik(x'-x'')}
\refstepcounter{equationA}
\eqno{(A\theequationA)}
\label{PPk}
$$
Using a regularisation similar to (\ref{fou1}), 
we integrate over $k$ and obtain
$$
  \langle\Phi_{x_1,b}(0)|\Phi_{x_0,b}(t)\rangle
    = \frac{1}{4\pi i b}\int_{x_1-b}^{x_1+b}\!\!\!dx'\int_{x_0-b}^{x_0+b}\!\!\!dx''
      \left(\frac{1}{t-x'+x''-i0}-\frac{1}{t+x'-x''-i0}\right)
\refstepcounter{equationA}
\eqno{(A\theequationA)}
\label{pp10}
$$
The integration over $x'$ and $x''$ and the rearrangement of terms give us
$$
  \langle\Phi_{x_1,b}(0)|\Phi_{x_0,b}(t)\rangle
    =  \psi_1(x_1-x_0+t-i0) - \psi_1(x_1-x_0-t+i0)
\refstepcounter{equationA}
\eqno{(A\theequationA)}
\label{pp11}
$$
where
$$
  \psi_1(x) 
    = \frac{1}{4\pi ib}\left((x-2b)\ln(x-2b) + (x+2b)\ln(x+2b) - 2x\ln(x)\right)
\refstepcounter{equationA}
\eqno{(A\theequationA)}
\label{psi1}
$$
Then, using (\ref{ln}), (\ref{arg}) we come to
$$ 
  \langle\Phi_{x_1,b}(0)|\Phi_{x_0,b}(t)\rangle 
    = \psi_2(x_1-x_0-t) + \psi_2^*(x_1-x_0+t)
\refstepcounter{equationA}
\eqno{(A\theequationA)}
\label{pp12}
$$
where
\begin{eqnarray}
  \psi_2(x) 
     & = & \frac{1}{8b}\left(|x-2b| +|x+2b| - 2|x|\right)  \nonumber \\
\label{psi2}
     & + & \frac{i}{4\pi b}\left(|x-2b|\,\ln|x-2b| + |x+2b|\,\ln|x+2b|
               - 2|x|\,\ln|x|\right) \nonumber
\end{eqnarray}
Fig. 5 shows the real  part, the imaginary part and the absolute value
of the scalar product $\langle\Phi_{x_1,b}(0)|\Phi_{x_0,b}(t)\rangle$ 
as functions of $t$ for two different values of the wave packet's width $b$.
The real component is non-vanishing only in the time interval
corresponding to the overlapping of the localized components.
In contrast, the imaginary part is non-vanishing immediately after $t=0$, because 
it reflects the overlapping of nonlocal components. 
Fig. 5 also shows that the ``causal'' part of the effect is much bigger
than the contribution of the long tails, if the size of the wave packet is much less
than the distance between the domains of localization.

\newpage
\listoffigures
\begin{tabular}{l l l}
1 & The real part (dashed lines) and the imaginary 
    part (solid lines)              & \\ 
  & of $\psi(x-t)$ (a),  $\psi^*(x+t)$ (c) 
    and $\Phi(x,t)=\psi(x-t)+\psi^*(x+t)$ (e)             & \\  
  & as functions of $x$ at $t = 0$;
    the absolute values $|\psi(x-t)|$ (b), $|\psi^*(x+t)|$ (d) & \\
  & and $|\Phi(x,t)| \neq |\psi(x-t)|+|\psi^*(x+t)|$ (f)
    as functions of $x$  at $t = 0$.\\
2 & The real part (dashed lines) and the imaginary 
    part (solid lines)              & \\ 
  & of $\psi(x-t)$ (a), $\psi^*(x+t)$ (c) 
    and $\Phi(x,t)=\psi(x-t)+\psi^*(x+t)$ (e)   & \\
  & as functions of $x$ at $t = 0.25\, s$; 
    the absolute values $|\psi(x-t)|$ (b), 
    $|\psi^*(x+t)|$ (d)              & \\      
  & and $|\Phi(x,t)| \neq |\psi(x-t)|+|\psi^*(x+t)|$ (f)  
    as functions of $x$  at $t = 0.25\, s$. \\
3 & The real part (dashed lines) and the imaginary 
    part (solid lines) of $\psi(x-t)$ (a),             & \\ 
  & $\psi^*(x+t)$ (c) and $\Phi(x,t)=\psi(x-t)+\psi^*(x+t)$ (e) 
    as functions of $x$                                 & \\
  & at $t = 1\, s$; the absolute values $|\psi(x-t)|$ (b), 
    $|\psi^*(x+t)|$ (d) and  & \\     
  & $|\Phi(x,t)| \neq |\psi(x-t)|+|\psi^*(x+t)|$ (f)  
    as functions of $x$  at $t = 1\, s$. \\
4 & $\langle\Phi_{0,b}(t)|\rho(x)|\Phi_{0,b}(t)\rangle$ 
    (evolving object) and  $\langle\Phi_{2,b}(0)|\rho(x)|\Phi_{2,b}(0)\rangle$ & \\
    & (right rectangle) with no overlap at $t=0$ (a), overlapping only by the  & \\
    & nonlocal tail at $t=0.75\, s$ (b), overlapping also by the local  (causal)   & \\
    & component at $t=1.25\, s$ (c).\\
5 & The real part (a) and (b), 
    the imaginary part (c) and (d),            
    the absolute value                                 & \\ 
  & (e) and (f) of 
    $\langle\Phi_{2,b}(0)|\Phi_{0,b}(t)\rangle$ as 
    functions of $t$ for $b=0.5\, ls$ ((a), (c), (e)) and   & \\
  &  $b=0.01\, ls$ ((b), (d), (f)).
\end{tabular}


\newpage
\begin{thebibliography}{99}

\bibitem{Heg98}
G. C. Hegerfeldt,
{\em Irreversibility and Causality in Quantum Theory - Semigroups and Rigged
Hilbert Spaces}, ed. by A. Bohm, H.-D. Dobner and P. Kielanovski, Springer Lecture Notes
{\bf 504}, (1998).

\bibitem{Fermi}
E. Fermi, Rev. Mod. Phys. {\bf 4}, 87 (1932).
\bibitem{Heg74}

G. C. Hegerfeldt,
{\em Phys. Rev.} {\bf D 10}, N 10, 3320-3321 (1974).

\bibitem{Heg79}
G. C. Hegerfeldt, S. N. Ruijsenaars
{\em Phys. Rev.} {\bf D 22}, N 2, 377-384 (1979).

\bibitem{Heg94}
G. C. Hegerfeldt,
{\em Phys. Rev. Lett.} {\bf 72}, N 5, 596-599 (1994).

\bibitem{NW}
T. D. Newton, E. P. Wigner,
{\em Reviews of Modern Physics} {\bf 21}, N 3, 400-406 (1949).

\bibitem{Pavlov}
"The positive subspace of the generator of the evolution
for the classical 1-d wave equation in the Energy normed space consists of
analytic functions from corresponding Hardy classes,
hence does not contain any localized data",
B.~S.~Pavlov, private communication.


\end{thebibliography}
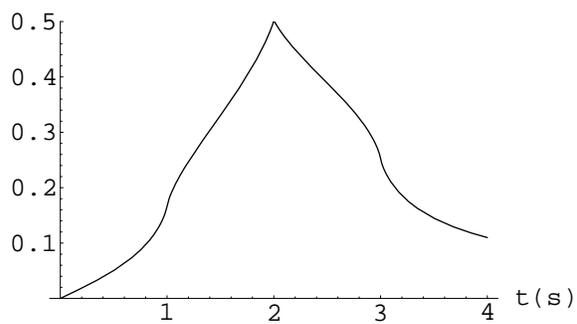
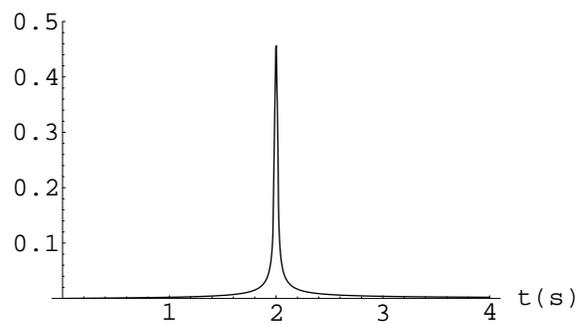
\end{document}